\documentclass[twocolumn,secnumarabic,amssymb, nobibnotes, aps, prl, superscriptaddress, nobalancelastpage,longbibliography,nofootinbib]{revtex4-1}
\usepackage{epsfig}
\usepackage{graphicx}
\usepackage{stmaryrd}
\usepackage{amssymb,tabularx,dcolumn}
\usepackage{supertabular,ltxtable}
\usepackage{longtable}
\usepackage{amsmath}
\usepackage{amstext}
\usepackage{float}
\usepackage{color}
\usepackage{bm}
\usepackage{hyperref}
\usepackage{lipsum}
\hypersetup{breaklinks=true,colorlinks=true,linkcolor=blue,citecolor=blue,filecolor=magenta,urlcolor=cyan}
\usepackage{mathtools}
\usepackage{multirow}
\usepackage{makecell}
\usepackage[utf8]{inputenc}
\usepackage{tabu}
\usepackage{braket}
\usepackage{subfigure}
\usepackage{enumitem}
\usepackage{orcidlink}
\usepackage{changes}

\usepackage{xcolor}
\definecolor{pastelgray}{rgb}{0.81, 0.81, 0.77}
\definecolor{beaublue}{rgb}{0.9, 0.9, 0.93}

\makeatletter
\def\@bibdataout@aps{%
\immediate\write\@bibdataout{%
@CONTROL{%
apsrev41Control%
\longbibliography@sw{%
    ,author="08",editor="1",pages="1",title="0",year="1"%
    }{%
    ,author="08",editor="1",pages="1",title="",year="1"%
    }%
  }%
}%
\if@filesw \immediate \write \@auxout {\string \citation {apsrev41Control}}\fi
}
\makeatother

\begin{document}

\preprint{APS/123-QED}

\title{Surprising  charge-radius kink in the Sc isotopes at \boldmath{$N$}=20}

\author{Kristian K\"onig\,\orcidlink{0000-0001-9415-3208}}
 \email{kkoenig@ikp.tu-darmstadt.de}
\affiliation{Facility for Rare Isotope Beams, Michigan State University, East Lansing 48824, USA}
\affiliation{Department of Nuclear Physics, Technische Universit\"at Darmstadt, 64289 Darmstadt, Germany}

\author{Stephan Fritzsche\,\orcidlink{0000-0003-3101-2824}}
\affiliation{Helmholtz Institute Jena, 07743 Jena, Germany}
\affiliation{Theoretisch-Physikalisches Institut, Friedrich-Schiller-Universit\"a{}t Jena, 07743 Jena, Germany}

\author{Gaute Hagen\,\orcidlink{0000-0001-6019-1687}}
\affiliation{Oak Ridge National Laboratory,
Oak Ridge, TN 37830, USA}
\affiliation{Department of Physics and Astronomy, University of
  Tennessee, Knoxville, Tennessee 37996, USA}

\author{Jason D. Holt\,\orcidlink{0000-0003-4833-7959}}
\affiliation{TRIUMF, Vancouver, British Colombia, V6T 2A3 Canada}
\affiliation{Department of Physics, McGill University, Montr\'eal, QC H3A 2T8, Canada}

\author{Andrew Klose}
\affiliation{Department of Chemistry and Biochemistry, Augustana University, Sioux Falls 57197, USA}

\author{Jeremy Lantis\,\orcidlink{0000-0002-8257-7852}}
\affiliation{Facility for Rare Isotope Beams, Michigan State University, East Lansing 48824, USA}
\affiliation{Department of Chemistry, Michigan State University, East Lansing 48824, USA}

\author{Yuan Liu}
\affiliation{Facility for Rare Isotope Beams, Michigan State University, East Lansing 48824, USA}

\author{Kei Minamisono\,\orcidlink{0000-0003-2315-5032}}
\email{minamiso@frib.msu.edu}
\affiliation{Facility for Rare Isotope Beams, Michigan State University, East Lansing 48824, USA}
\affiliation{Department of Physics and Astronomy, Michigan State University, East Lansing 48824, USA}

\author{Takayuki Miyagi\,\orcidlink{0000-0002-6529-4164}}
\affiliation{Department of Nuclear Physics, Technische Universit\"at Darmstadt, 64289 Darmstadt, Germany}
\affiliation{ExtreMe Matter Institute EMMI, GSI Helmholtzzentrum f\"ur Schwerionenforschung GmbH, 64291 Darmstadt, Germany}
\affiliation{Max-Planck-Institut f\"ur Kernphysik, Saupfercheckweg 1, 69117 Heidelberg, Germany}

\author{Witold Nazarewicz\,\orcidlink{0000-0002-8084-7425}}
\affiliation{Facility for Rare Isotope Beams, Michigan State University, East Lansing 48824, USA}
\affiliation{Department of Physics and Astronomy, Michigan State University, East Lansing 48824, USA}


\author{Thomas~Papenbrock\,\orcidlink{0000-0001-8733-2849}}
\affiliation{Department of Physics and Astronomy, University of
  Tennessee, Knoxville, Tennessee 37996, USA}
\affiliation{Physics Division, Oak Ridge National Laboratory, Oak
  Ridge, Tennessee 37831, USA}

\author{Skyy V. Pineda\,\orcidlink{0000-0003-1714-4628}}
\affiliation{Facility for Rare Isotope Beams, Michigan State University, East Lansing 48824, USA}
\affiliation{Department of Chemistry, Michigan State University, East Lansing 48824, USA}

\author{Robert Powel}
\affiliation{Facility for Rare Isotope Beams, Michigan State University, East Lansing 48824, USA}
\affiliation{Department of Physics and Astronomy, Michigan State University, East Lansing 48824, USA}

\author{Paul-Gerhard Reinhard\,\orcidlink{0000-0002-4505-1552}}
\affiliation{Friedrich-Alexander Universität, Erlangen 91058, Germany}

\date{\today}

\begin{abstract}
Charge radii of neutron deficient $^{40}$Sc and $^{41}$Sc nuclei were determined using collinear laser spectroscopy. With the new data, the chain of Sc charge radii extends below the neutron magic number $N = 20$ and shows a pronounced kink, generally taken as a signature of a shell closure, but one notably absent in the neighboring Ca, K and Ar isotopic chains. Theoretical models  that explain the trend at $N = 20$ for the Ca isotopes cannot reproduce this puzzling behavior.  
\end{abstract}

\maketitle

\paragraph{Introduction:} 
The introduction of the nuclear shell model \cite{Goeppert.1949, Haxel.1949} enabled the understanding of diverse observables, such as nuclear global properties and  excitation energies. The numbers of nucleons that completely fill a shell are known as magic numbers, which are 2, 8, 20, 28, 50, 82, and 126 in stable nuclei. The simple structure of magic nuclei facilitate comparison with theories. With the advent of radioactive beam facilities, short-lived nuclei far from stability can be investigated, in which new magic numbers are found and the traditional shell closures vanish \cite{Sorlin.2008}.
Particularly, the Ca isotopic chain at the proton-shell closure  $Z=20$ has been of great interest for nuclear structure studies since it features two stable doubly-magic isotopes $^{40,48}$Ca.  The intricate pattern of charge radii along the Ca chain \cite{Pendrill.1992, Vermeeren.1992, Garcia.2016,Vermeeren.1996, Miller.2019} has been a long-standing challenge for  many-body nuclear theory, see discussion in, e.g.,  Refs.~\cite{Garcia.2016,Miller.2019,Kortelainen2022}. 

One interesting feature around magic gaps is a discontinuity, or a kink, in charge radii along isotopic or isotonic chains \cite{Otten.1989, Campbell.2016}. A number of theoretical models have been proposed to explain the kink, including pairing correlations, particle-vibrational coupling and ground-state vibrational correlations, and spin-orbit effects \cite{Gorges2019,Reinhard.2021,Perera.2021}. The relation between the robustness of the magic number and the magnitude of the kink, however, is not obvious. In the Ca chain at $N = 28$, there is a prominent kink, whose magnitude is comparable with that of the doubly-magic $^{56}$Ni \cite{Sommer.2022}, which is known to be a soft  doubly-magic nucleus \cite{ots98} prone to polarization effects.

There is one exception at the magic number $N =20$ for the Ca isotopic chain \cite{Vermeeren.1996, Miller.2019}, where only a smooth variation of charge radii has been observed.  The absence of a kink extends to the neighboring Ar \cite{Klein.1996} and K \cite{Touchard.1982, Behr.1997, Rossi.2015} chains. The origin of the apparent disappearance of kinks in the Ar, K and Ca chains is still an open question. To shed light on this unusual behavior, we carried out a measurement of the charge radii across the $N=20$ shell closure for the neutron-deficient Sc isotopes with $Z=21$. At low energies, the Sc nuclei have an additional proton in the $0f_{7/2}$ shell outside the magic Ca core. The presence of the odd proton strongly impacts the core polarization as evidenced by strong shape coexistence effects in the stable $^{45}$Sc isotope \cite{Bednarczyk1997}. The single proton in the $0f_{7/2}$ shell in $^{45}$Sc may couple to the spherical $^{44}$Ca core. On the other hand, a proton hole in the $0d_{3/2}$ shell weakly couples to the deformed $^{46}$Ti core and leads to a collective behavior. Since $^{40}$Sc lies  at the proton dripline \cite{Woods.1988},  continuum effects may impact the charge radius \cite{Miller.2019}, though due to the blocking effect of the unpaired neutron and proton  in the $0d_{3/2}$ and $0f_{7/2}$ shell, respectively, the pairing effects are   expected to be suppressed.

\paragraph{Experiment:}
The $^{40, 41}$Sc isotopes were produced at the National Superconducting Cyclotron Laboratory by nucleon pick-up of a $^{40}$Ca primary beam at 140 MeV/nucleon in a Be target. The ions were separated using the A1900 fragment separator \cite{Morrissey.2003} and transported to a gas stopper \cite{Cooper.2014}, where the fast beam was thermalized. At a beam energy of 30 keV, the singly-charged bare ions were transported to the BECOLA facility \cite{Minamisono.2013, Rossi.2014} with rates of approximately 15\,000 and 20\,000 ions/s for $^{40}$Sc$^+$ and $^{41}$Sc$^+$, respectively.  
There, the ions were guided into a helium-gas filled radio-frequency quadrupole trap (RFQ) \cite{Barquest.2017}, where they were cooled, accumulated and extracted as ion bunches to reduce the laser background \cite{Nieminen.2002}. The ion beam was overlapped with a collinear laser beam using a 30$^\circ$ electrostatic deflector. Between two 3-mm apertures placed 2.1\,m apart, photons were counted with photo-multiplier tubes that were installed on top of a mirror-based detection system \cite{Minamisono.2013, Maass.2020}. 
This detection region was floated on a scanning potential to perform Doppler tuning, which allowed us to operate the laser at a constant frequency while varying this potential by 250\,V.
To reference the isotope shift and to monitor potential long-term drifts, stable $^{45}$Sc$^+$ was introduced into the BECOLA RFQ from an offline Penning-ionization-gauge source \cite{Ryder.2015} and probed every 4-6\,h. When switching between the isotopes, the beam energy was kept constant and the laser frequency was adapted to match the different Doppler shift.

The $3p^63d4s\;^3D_2 \rightarrow 3p^63d4p\;^3F_3$ at 363.2\,nm transition was investigated in $^{40,41,45}$Sc$^+$. 
The employed continuous-wave Ti-sapphire laser (Matisse TS, Sirah Lasertechnik) was stabilized to a wavemeter (WSU30, HighFinesse) and operated at 726\,nm.  
The light was frequency doubled (Wavetrain, Spectra Physics) to 363\,nm, transported via optical fiber to the beamline and irradiated in collinear geometry. Spectroscopy was performed with a laser power of 300\,$\mu$W and a laser-beam diameter of 1\,mm.

\begin{figure}
	\centering
		\includegraphics[width=0.4500\textwidth]{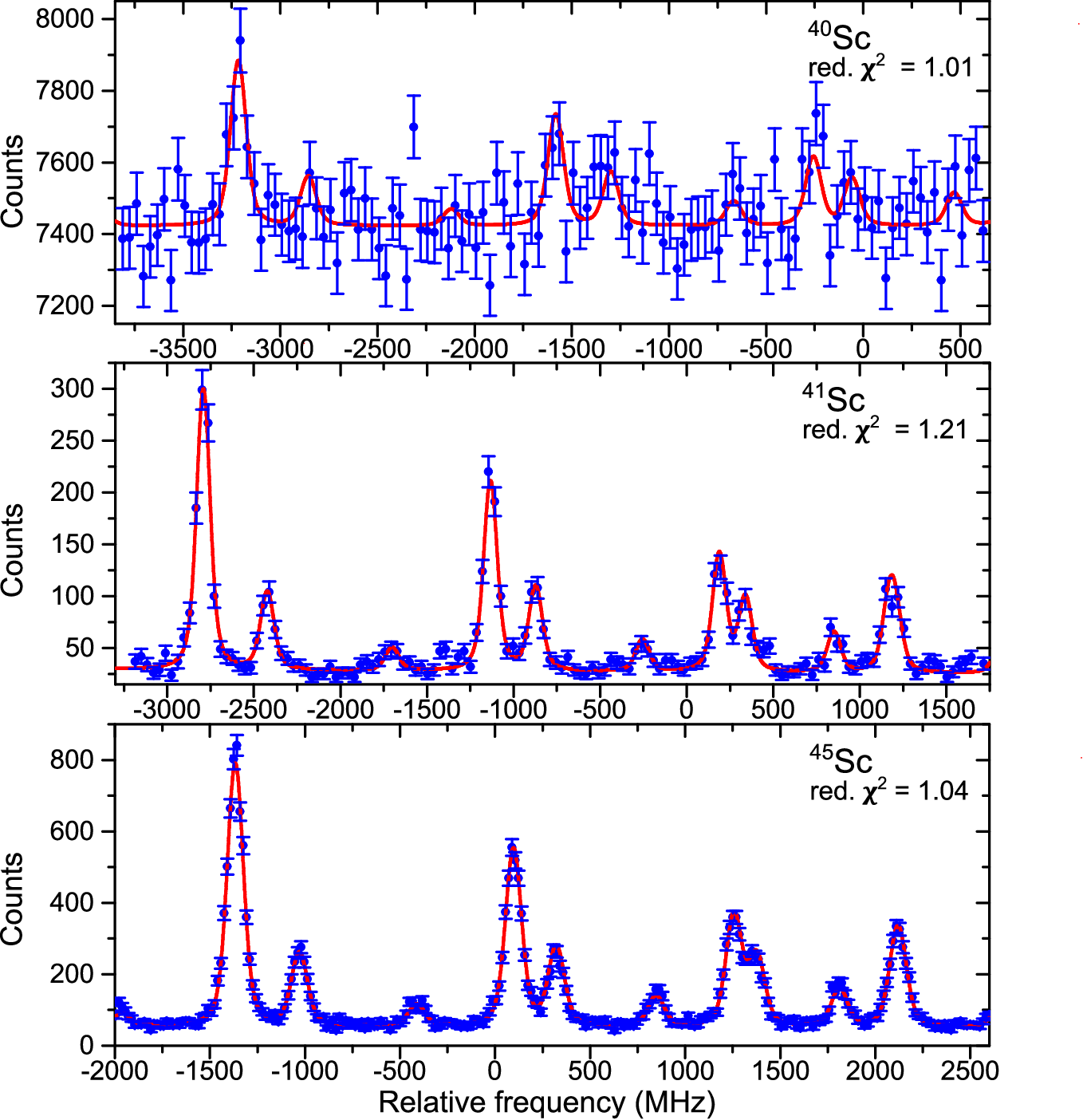}
	\caption{Resonance spectra of the $3p^63d4s\;^3D_2 \rightarrow 3p^63d4p\;^3F_3$ transition in $^{40,41,45}$Sc. The frequency is relative to the transition center of gravity of $^{45}$Sc.
	}
	\label{fig:Spectra}
\end{figure}

\begin{table*}
\caption{\label{tab:IsoShift} Isotope shifts $\delta\nu^{A,A'}$, differential mean square charge radii $\delta \langle r^2\rangle^{A,A'}$ and rms charge radii $R_\mathrm{ch}$. The isotope shifts of $^{42-46}$Sc are from \cite{Avgoulea.2011,Koszorus.2021b} but the nuclear charge radius was updated with the improved calculations of the atomic factors. The statistical and systematic uncertainty is given in parentheses. In the systematic uncertainty of $R$ the uncertainty of the reference isotope $^{45}$Sc of 0.0025\,fm \cite{Angeli.2013} is included. }
\begin{ruledtabular}
\begin{tabular}{c c c c c c}
$A, A'$ & $I^P(A)$ & $\delta\nu^{A,A'}$ & Reference & $\delta \langle r^2\rangle^{A,A'}$ & $R_\mathrm{ch}(^{A}$Sc) \\
    & &  (MHz)   &       &  (fm$^2)$          &  (fm) \\
\hline
40, 45  & 4$^-$   & -1594 (8)  & This work & -0.226 (22) (175)	& 3.514	(3) (25) \\
41, 45  & $7/2^-$ & -1199 (4)  & This work &  -0.305 (10) (137) & 3.503 (1) (20) \\
42, 45  & $0^+$   & -985 (11)    & \cite{Avgoulea.2011} & 0.076	(31) (100)	& 3.557	(4) (14) \\
42m, 42 & $7^+$   &  +74 (5)      & \cite{Koszorus.2021b} & -0.210\,(14)\,(7)  & 3.527\,(5)\,(14) \\
43, 45  & $7/2^-$ & -631 (5)     & \cite{Avgoulea.2011} & 0.019(14)(65)	& 3.549	(2)	(10) \\
44, 45  & $2^+$   & -287 (4)     & \cite{Avgoulea.2011} &  -0.051\,(11)\,(32)  &  3.539\,(2)\,(5) \\
44m, 44 & $6^+$   & +25 (4)      & \cite{Avgoulea.2011} & -0.071\,(11)\,(2)  & 3.529\,(2)\,(5) \\
45, 45  & $7/2^-$ & 0           & \cite{Angeli.2013} &  0  &  3.5459\,(0)\,(25) \\
45m, 45 & $3/2^+$ & -66 (2)  &\cite{Avgoulea.2011} & 0.187\,(6)\,(6) & 3.572\,(1)\,(3) \\
46, 45  & $4^+$   & 336 (3)  & \cite{Avgoulea.2011} &  -0.124\,(9)\,(31)  &  3.528\,(1)\,(5) \\
\end{tabular}
\end{ruledtabular}
\end{table*}

\paragraph{Results:}
The obtained spectra are shown in Fig.\,\ref{fig:Spectra}.
The lower signal quality of the $^{40}$Sc$^+$ resonance spectrum is caused by a significant contamination with $^{40}$Ar$^+$ from the gas stopping cell. Its rate exceeded the $^{40}$Sc$^+$ rate by three orders of magnitude.
To avoid overfilling the BECOLA RFQ, a short accumulation time of $t_\mathrm{acc}(^{40}\mathrm{Sc}^+)=20$\,ms had to be chosen, leading to a less efficient background suppression compared to $^{41}\mathrm{Sc}^+$ with $t_\mathrm{acc}(^{41}\mathrm{Sc}^+)=1$\,s.

The deduced isotope shifts $\delta \nu^{A,45}= \nu^A-\nu^{45}$ are listed in Table\,\ref{tab:IsoShift} together with literature values of other Sc isotopes. The statistical uncertainty was 1.7 and 6.8\,MHz for $^{41}$Sc$^+$ and $^{40}$Sc$^+$, respectively, and approximately 1.1\,MHz for the $^{45}$Sc$^+$ reference spectra.
The kinetic beam energy was determined with $10^{-5}$ relative accuracy by evaluating the $^{45}$Sc$^+$ resonance frequencies together with its rest-frame frequency, which was measured via collinear and anticollinear spectroscopy beforehand as described in Ref. \cite{Koenig.2021}, leading to a 0.2-MHz uncertainty contribution.

To consider a varying filling rate of the buncher due to different purity of the ion beam, a general 2\,MHz uncertainty is included, which corresponds to observations from stable beam measurements at BECOLA \cite{Koenig.2021b}. The laser frequency measurement was realized with a wavemeter. 
According to a detailed study on devices from the same manufacturer \cite{Verlinde.2020, Koenig.2020} that we confirmed for our device \cite{Koenig.2021b}, a 2.8-MHz contribution was considered.
Due to the poor signal-to-noise ratio, the $^{40}$Sc$^+$ spectrum could not be fitted without constraints and the ratios of the $A$ and $B$ parameters between the upper and lower level were fixed to those of $^{45}$Sc$^+$. Nevertheless, the uncertainty of the $B$ parameter was large and differed by $1.2\sigma$ from the theoretically expected value ($B_\mathrm{lower}(\mathrm{exp.})=58(64)$\,MHz, $B_\mathrm{lower}(\mathrm{theo.})=-19$\,MHz), see Ref. \cite{Powel.2022} for a detailed study of the hyperfine splitting.
Fixing the $B$ parameter to the theoretical value leads to a shift of the center frequency by 2.3\,MHz, which we consider as additional uncertainty. Adding the described contributions in quadrature, leads to a total uncertainty of 8.1 and 4.0\,MHz for $^{40}$Sc and $^{41}$Sc, respectively.




\paragraph{Atomic Computations:}
To describe the electronic response of the charge radii, detailed computations need to be performed not only for the level structure but also the isotope (and hyperfine) parameters. Here, we have applied the multi-configuration Dirac-Hartree-Fock (MCDHF) method \cite{Grant.2007}, based on different advanced models, to generate the state functions and the electronic mass-shift $K^\mathrm{(MS)}$ and field-shift parameter $F^\mathrm{\,(el)}$ for the standard parametrization of atomic isotope shifts. For light elements, such as Sc, the mass shift $K^\mathrm{(MS)} \:=\: K^\mathrm{(NMS)} \:+\: K^\mathrm{(SMS)}$, with $K^\mathrm{(NMS)}$ being the so-called \textit{normal} and $K^\mathrm{(SMS)}$ the \textit{specific} mass-shift coefficient, is known as most critical. Since these mass-shift parameters are very sensitive to electron-electron correlations, the active space method with sizable expansions and virtual -- single, double, and partly triple -- excitations  into additional layers of correlation orbitals need generally to be applied for their computations. Despite the advances in atomic theory during recent years, the nearly degenerate $3d$, $4s$ and $4p$ shells require to open also the $3s$ and $3p$ shells for their core-core contributions. This has made accurate isotope-shift parameters for open $3d$-shell elements a great challenge until today.

Three series of computations have been performed during the past decade for the $3p^6 3d 4s\;\, ^3D_2 \rightarrow 3p^6 3d 4p\;\, ^3F_3$ transition, based on the MCDHF method. These series are based on the independent implementations of this method with the RATIP \cite{Fritzsche.2012}, GRASP \cite{Jonsson.2013,Naze.2013}, and JAC \cite{Fritzsche.2019} codes
and give rise to the isotope shift parameters in Table\,\ref{tab:parameters}.
The (numerical) uncertainties reflect the overall stability of the computations by using a separate optimization of the upper and lower levels. To ensure their balance in the calculations, all series were  started from a frozen set of occupied (spectroscopic) orbitals. 
The shear size of the computations require to apply the weighted-mean value and the standard deviation to extract the (total) mass-shift parameter. 
Although these practical arguments cannot exclude a systematic shift, for instance due to the omission of relevant interaction and correlation contributions, the independent set-up of the codes and computations reduce this risk considerably. In the analysis below, we use $K^\mathrm{(MS)}=604 \pm  22$\,GHz\,u and $F^\mathrm{\,(el)}=-352 \pm 12$\,MHz/fm$^2$.

The resulting differential mean square charge radii $\delta \langle r^2 \rangle^{A,45}$ relative to the stable $^{45}$Sc as well as the root-mean-square (rms) charge radii $R_\mathrm{ch}$ are listed in Table\,\ref{tab:IsoShift}. Note that the updated mass- and field-shift parameters were also applied to the previous measurements \cite{Avgoulea.2011, Koszorus.2021b}, which led to consistent charge radii, but smaller uncertainties.

\begin{table}
\caption{\label{tab:parameters} Comparison of the total mass-shift $K^\mathrm{(MS)}$ and field-shift $F^\mathrm{\,(el)}$ parameters as obtained from three independent implementations and series of computations.}
\begin{ruledtabular}
\begin{tabular}{ccc}
    Series   &  $K^\mathrm{(MS)}$ & $F^\mathrm{(el)}$   \\ 
     & (GHz~u) & (MHz/fm$^2$) \\
      \colrule
    Ref. \cite{Avgoulea.2011}  &  $+\,583 \pm  30$          &       $-\,355 \pm 50 $                \\ 
    Ref. \cite{Koszorus.2021b}  &  $+\,625 \pm  60$         &       $-\,349 \pm 15 $                \\ 
    This work           &  $+\,633 \pm  40$          &       $-\,358 \pm 20 $                \\ 
    Weighted mean           &  $+\,604 \pm  22$          &       $-\,352 \pm 12 $                \\ 
\end{tabular}
\end{ruledtabular}
\end{table}

\begin{figure*}
	\centering
		\includegraphics[width=0.68\textwidth]{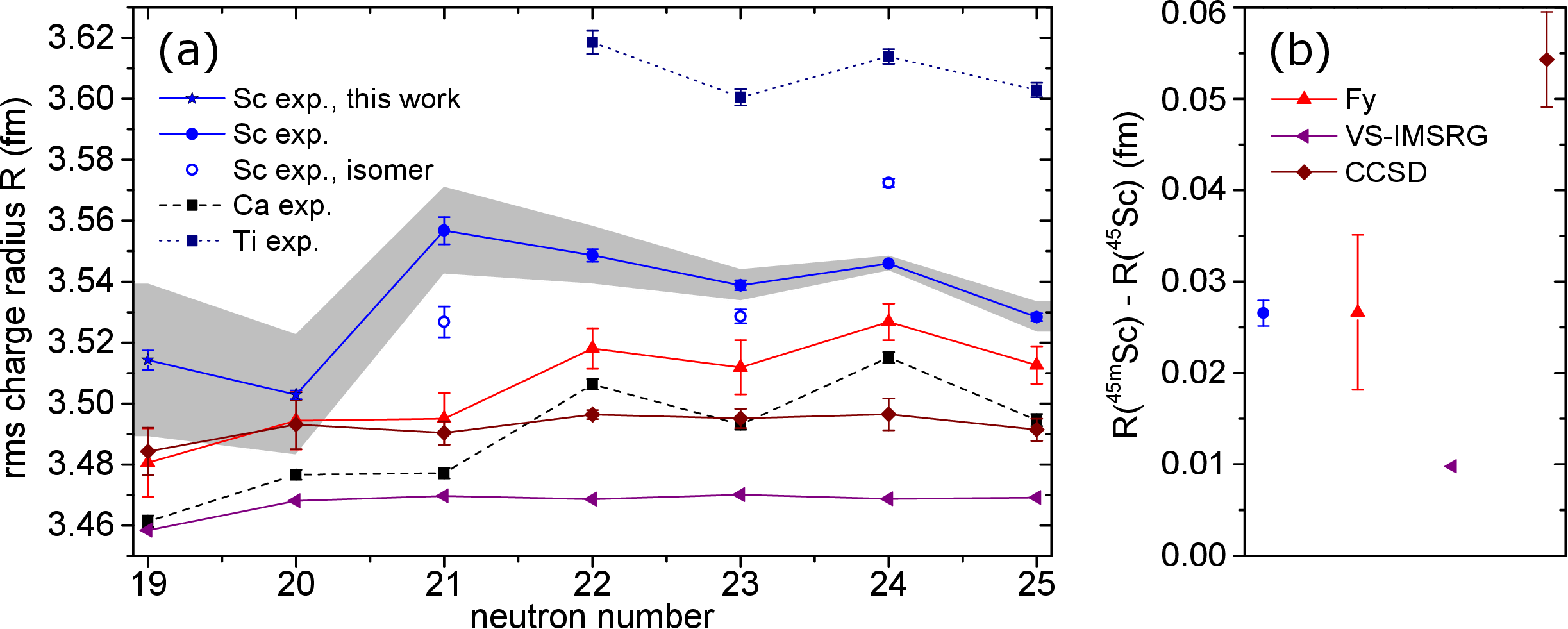}
	\caption{(a): Experimental and theoretical rms charge radii of Sc. The error bars indicate the experimental uncertainty while the gray shaded area is the uncertainty that originates in the calculated mass and field shift factors. The atomic factors can cause a different tilt along the isotopic chain but do not affect the overall trend, i.e., the appearance of a kink at $N=20$. The experimental radii of isomeric Sc states \cite{Avgoulea.2011, Koszorus.2021b} are plotted with open circles and, as a reference, the radii of the Ca \cite{Miller.2019} and Ti \cite{Gangrsky.2004} isotopic chains are depicted.
    (b): Measured and calculated charge radius difference between the $3/2^+$ isomeric state and the $7/2^-$ ground state of $^{45}$Sc. When calculating the  uncertainty of the Fy result, it was assumed that the calculated radii are not correlated; hence the corresponding error bar 
    should be viewed as an upper limit \cite{ReinWit2022}.
	}
	\label{fig:ScRadii}
\end{figure*}

\paragraph{Discussion:}
The measured rms charge radii of Sc isotopes are plotted in Fig.\,\ref{fig:ScRadii} as  stars (this work) and circles (results from Refs.~\cite{Avgoulea.2011, Koszorus.2021b}). The error bars correspond to the experimental uncertainty, while the grey area shows the full uncertainty, which is dominated by the uncertainty of the calculated mass- and field-shift parameters. In $^{42,44,45}$Sc, there are isomeric states \cite{Avgoulea.2011, Koszorus.2021b}, whose charge radii are also shown in Fig.\,\ref{fig:ScRadii}.  
For $N\geq22$, the ground state  charge radii exhibit a similar trend to what has been observed for the Ca  and Ti chains. Significant differences, however, are seen in the neutron-deficient isotopes. In particular, the rms charge radius of $^{41}$Sc is significantly below that of $^{42}$Sc, and the charge radius of $^{40}$Sc rises with respect to that for $^{41}$Sc resulting in a pronounced kink structure at $N = 20$. The relatively large systematic uncertainty due to the atomic calculations cannot inhibit the kink since the variation of atomic factors can only tilt the entire trend around the $^{45}$Sc pivot point. 


The charge radii for the Sc isotopes were computed with the nuclear density functional (DFT), the coupled-cluster, and the valence-space in-medium similarity renormalization group (VS-IMRG) theory. In DFT, we employed the functional Fy($\Delta r$, HFB)  \cite{Miller.2019} that contains novel gradient terms in pairing and surface energies. Its parameters have been calibrated to the  large dataset of ground state properties in semi-magic nuclei \cite{Kluepfel.2009}. 
This model has been particularly successful in describing the neighboring Ca, K and Ti isotopic chains \cite{Miller.2019,Koszorus.2021, Kortelainen2022}. 
The DFT calculations were done with the axial HFB solver which allows for deformation and spin polarization. In order to compute the low-lying one- and two-quasi-particle excitations, we blocked all quasi-particle HFB states in the energy window of $\pm 6$\,MeV around the Fermi level. For even-$N$ systems, we only blocked the proton quasi-particles while for odd-$N$ systems we blocked both protons and neutrons. The resulting total energies were sorted energetically. All computed low-energy configurations turned out nearly spherical (within quadrupole deformation $|\beta_2| < 0.06$) and energies 
bunching approximately according to underlying spherical shells. As expected, the $0f_{7/2}$ shell corresponds to the lowest excitations for protons and for neutrons with $N > 20$. For $N = 19$, the $0d_{3/2}$  neutron shell yields the lowest energies. For even-$N$ systems, the lowest configuration  has $I^\pi = 7/2^-$ and we predict $I^\pi=3/2^+$ isomers based on the $0d_{3/2}\rightarrow 0f_{7/2}$ proton excitation. These isomers are prolate-deformed with $\beta_2\approx 0.25$, which agrees with the mean-field predictions of Ref.~\cite{Bednarczyk1997}. 
For odd-$N$ isotopes, the situation is somehow ambiguous because we do not carry out the  angular momentum 
and isospin projection, which is essential when approaching $N\approx Z$ nuclei
\cite{Satula2012,Satula2016}. Thus, for given angular momentum $I$, we averaged over all combinations with $I = I_p + I_n$. This average value is taken as an estimate of the angular momentum and the resulting variance serves as an estimate of the systematic error of the approximate projection. The second source of uncertainty is the statistical resulting from the parameter calibration. 
Calculations with broken spherical symmetry and broken time reversal symmetry  result in changes of charge radii amounting to  up to 0.01\,fm.

The coupled-cluster  and VS-IMSRG calculations employ the $\Delta$NNLO$_\text{GO}(394)$ chiral nucleon-nucleon and three-nucleon interaction with explicit Delta isobars~\cite{jiang2020}. The coupled-cluster calculations~\cite{kuemmel1978,bartlett2007,hagen2014}  start from an axially symmetric Hartree-Fock reference state built from 13 spherical major oscillator shells with the oscillator frequency $\hbar\omega = 16$~MeV. The three-nucleon force had an additional energy cut of $E_{\mathrm 3max} = 16 \hbar\omega$. We consider two different types of Hartree-Fock reference states. The neutron(s) at the Fermi surface are in the $0f_{7/2}$ shell and fill pairs with $\pm j_z$, starting with $|j_z|=1/2$; an odd neutron is then in the minimum positive $j_z$ state. The proton in the $0f_{7/2}$ shell either has angular momentum projection $j_z = 1/2$ or $j_z=7/2$. The difference between both occupations is marked as an uncertainty.  
The former filling yields prolate deformations and the tentative spin-parity assignments are $I^\pi=1/2^-, 1^+, 1/2^-, 2^+,1/2^-, 3^+$ for $^{41,42,43,44,45}$Sc, respectively, while the latter tentatively yields high spins and is more consistent with experiment. For $^{40}$Sc we have the odd neutron in the $0d_{3/2}$ shell with $j_z=3/2$ and the spin-parity assignment is $I^\pi=2^-$. Here, we assumed that $I=|I_z|$ as we did not perform the angular-momentum projection. For the $3/2^+$ isomer in $^{45}$Sc we had a proton hole on the $j_z=3/2$ orbital of the $0d_{3/2}$ shell and two protons with $j_z=\pm 1/2$ in the $0f_{7/2}$ shell. 
The coupled-cluster computations are performed in the singles- and doubles approximation (CCSD). The point-proton radii are computed as an expectation value using the left and right ground state of the similarity transformed coupled-cluster Hamiltonian~\cite{novario2020}. To obtain the charge radii we include relativistic corrections and nucleon finite size effects~\cite{hagen2015}.   

The valence-space in-medium similarity renormalization group (VS-IMSRG)~\cite{Stro17ENO,Stroberg2019} is used to construct an approximate unitary transformation to decouple a multi-shell valence-space Hamiltonian~\cite{Miyagi2020} for proton-neutron $\{1s_{1/2}, 0d_{3/2}, 0f_{7/2}, 1p_{3/2}\}$ orbits above a $^{28}$Si core, allowing for excitation across the $Z=N=20$ shell gap.
The VS-IMSRG decoupling is done within the 13 spherical major oscillator space with the frequency of 16 MeV.
The 3N interaction matrix elements are included up to the sufficiently large truncation $E_{\rm 3max}=24 \hbar\omega$~\cite{Miyagi2022}.
The exact diagonalizations within the valence-space are performed with the KSHELL code~\cite{Shimizu2019}.

As illustrated in Fig.\,\ref{fig:ScRadii}, the evolution of charge radii for $N\geq 22$ is well described by the Fy functional, which predicts a similar trend as in  the Ca isotopes. The VS-IMSRG and CCSD calculations systematically 
underestimate the charge radii along the Sc chain.
At the neutron-deficient side, all employed models are unable to explain the experimental trend below $N=22$. In particular, they fail in reproducing the drastic decrease of charge radii between $N=21$ and $N=20$. 

In the heavier odd-even system of $^{45}$Sc, the charge radius difference $\Delta R \equiv R(3/2^+)-R(7/2^-)$ between the deformed isomer and ground state, shown in Fig.\,\ref{fig:ScRadii}b,   is correctly predicted by Fy, by taking into account quadrupole polarization effect of the proton $0d_{3/2}\rightarrow 0f_{7/2}$ excitation \cite{Bednarczyk1997}. 
Indeed, the weak coupling of the proton hole in the $0d_{3/2}$ shell to the deformed $^{46}$Ti core causes the increase of the radius while the single proton in the $0f_{7/2}$ shell in $^{45}$Sc  couples to the spherical $^{44}$Ca core  \cite{Bednarczyk1997}. As compared to experiment and Fy calculations, the VS-IMSRG approach significantly underestimates $\Delta R(^{45}$Sc), which is most likely due to its restricted configuration space  resulting in the under-predicted quadrupole collectivity  \cite{Stroberg2022}. In contrast, the CCSD approach employing the axially deformed  Hartree-Fock reference state 
is capable of exploring larger collective spaces; this results in a fairly large value of $\Delta R$ and systematically larger values of charge radii along the Sc chain are predicted by this approach as compared to VS-IMSRG, see Fig.\,\ref{fig:ScRadii}a.

The positive parities and smaller radii of isomeric states in  $^{42m}$Sc and $^{44m}$Sc suggest that these  excitations involve $0f_{7/2}$ neutrons and  $0f_{7/2}$ protons that do not lead to strong polarization effects. The low values of charge radii in $^{42m}$Sc and $^{44m}$Sc    have not yet been explained \cite{Avgoulea.2011, Koszorus.2021b}. It has to be stressed that the $T=0$, $I=0$ ground state of  $^{42}$Sc
cannot be  represented by a single mean-field configuration \cite{Satula2012}. This suggests \cite{Satula2016} that when it comes to the DFT description of charge radii differences between the isovector and isoscalar configurations of $^{42}$Sc and $^{44}$Sc, a multi-reference approach involving angular-momentum and isospin projection is required.

\paragraph{Summary}
Charge radii of the neutron deficient scandium isotopes $^{40}$Sc and $^{41}$Sc were determined using collinear laser spectroscopy. The new data demonstrate the presence of an appreciable kink structure at the $N$ = 20 neutron shell closure, which is absent in the neighboring Ca, K, and Ar isotopic chains.

The experimental data on charge radii were interpreted using ab-initio and DFT models employing realistic interactions. All models are consistent with the data for $N\ge 22$ but fail in reproducing the experimental pattern in charge radii for neutron-deficient isotopes, including the kink structure at $N=20$. At the same time, the employed models have been successful in explaining the absence of the kink in the  Ca isotopic chain. We thus conclude that the charge radius anomaly at $^{41}$Sc poses a significant puzzle for nuclear theory. Additional measurements of charge radii in the Ti isotopic chain across $N$ = 20 are called for as well as further theoretical studies of core polarization effects in this mass region.

\begin{acknowledgments}
\paragraph{Acknowledgements}
We thank S. R. Stroberg for the imsrg++ code~\cite{Stro17imsrg++} used to perform VS-IMSRG calculations. This work was supported in part by the National Science Foundation, Grants No. PHY-15-65546 and No. PHY-21-11185; by the U.S. Department of Energy under Award Numbers  DE-SC0013365 (Office of Science), DE-SC0023175 (Office of Science, NUCLEI SciDAC-5 collaboration), DE-FG02-96ER40963, DE-SC0018223, and under contract DE-AC05-00OR22725 with UT-Battelle, LLC (Oak Ridge National Laboratory), by NSERC under grants SAPIN-2018-00027 and RGPAS-2018-52245, the Arthur B. McDonald Canadian Astroparticle Physics Research Institute, the Deutsche Forschungsgemeinschaft (DFG, German Research Foundation) -- Project-ID 279384907 -- SFB 1245, and the European Research Council (ERC) under the European Union’s Horizon 2020 research and innovation programme (Grant Agreement No.\ 101020842). TRIUMF receives funding via a contribution through the National Research Council of Canada.This project has received funding from the European Union’s Horizon 2020 research and innovation program under grant agreement No 861198-LISA-H2020-MSCA-ITN-2019. Computer time was provided by the Innovative and Novel Computational Impact on Theory and Experiment (INCITE) program. This research used resources of the Oak Ridge Leadership Computing Facility at the Oak Ridge National Laboratory, which is supported by the Office of Science of the U.S. Department of Energy under Contract No. DE-AC05-00OR22725. VS-IMSRG omputations were performed with an allocation of computing resources on Cedar at WestGrid and Compute Canada.
\end{acknowledgments}

\bibliography{literature}
\end{document}